\documentclass[letterpaper]{article} 
\usepackage{aaai2026}  
\usepackage{times}  
\usepackage{helvet}  
\usepackage{courier}  
\usepackage[hyphens]{url}  
\usepackage{graphicx} 
\usepackage{natbib}  
\usepackage{caption} 
\usepackage{placeins}

\usepackage[percent]{overpic}
\usepackage{amsmath}
\usepackage{amsthm}
\usepackage{amsfonts}
\usepackage{booktabs}
\usepackage{xcolor}

\nocopyright
\title{Towards Scalable Proteomics: Opportunistic SMC Samplers on HTCondor}
\author {
    Matthew Carter\textsuperscript{\rm 1},
    Lee Devlin\textsuperscript{\rm 1},
    Alexander Philips\textsuperscript{\rm 1},
    Edward Pyzer-Knapp\textsuperscript{\rm 3},\\
    Paul Spirakis\textsuperscript{\rm 2} and
    Simon Maskell\textsuperscript{\rm 1}
}
\affiliations {
    \textsuperscript{\rm 1}Department of Electrical Engineering and Electronics, University of Liverpool, United Kingdom\\
    \textsuperscript{\rm 2}Department of Computer Science, University of Liverpool, United Kingdom\\
    \textsuperscript{\rm 3}Xyme, Manchester, United Kingdom\\
    matthew.carter@liverpool.ac.uk, a.m.philips@liveprool.ac.uk, lee.devlin@liverpool.ac.uk, ed@xyme.ai, P.Spirakis@liverpool.ac.uk, smaskell@liverpool.ac.uk
}

\begin{document}

\maketitle

\begin{abstract}
Quantitative proteomics plays a central role in uncovering regulatory mechanisms, identifying disease biomarkers, and guiding the development of precision therapies. These insights are often obtained through complex Bayesian models, whose inference procedures are computationally intensive, especially when applied at scale to biological datasets. This limits the accessibility of advanced modelling techniques needed to fully exploit proteomics data. Although Sequential Monte Carlo (SMC) methods offer a parallelisable alternative to traditional Markov Chain Monte Carlo, their high-performance implementations often rely on specialised hardware, increasing both financial and energy costs. We address these challenges by introducing an opportunistic computing framework for SMC samplers, tailored to the demands of large-scale proteomics inference. Our approach leverages idle compute resources at the University of Liverpool via HTCondor, enabling scalable Bayesian inference without dedicated high-performance computing infrastructure. Central to this framework is a novel Coordinator-Manager-Follower architecture that reduces synchronisation overhead and supports robust operation in heterogeneous, unreliable environments. We evaluate the framework on a realistic proteomics model and show that opportunistic SMC delivers accurate inference with weak scaling, increasing samples generated under a fixed time budget as more resources join. To support adoption, we release CondorSMC, an open-source package for deploying SMC samplers in opportunistic computing environments.
\end{abstract}

\begin{links}
    \link{Code}{https://github.com/UoL-SignalProcessingGroup/CondorSMC}
\end{links}

\section{Introduction}
\label{sec:introduction}
Quantitative proteomics, the large-scale study of protein expression and abundance, plays a critical role in understanding biological processes and disease mechanisms. Bayesian models are increasingly used in this domain to capture uncertainty, handle missing data, and integrate multiple sources of evidence. However, their application is often constrained by computational demands, particularly for high-dimensional or hierarchical models \cite{proteinquantification}. As datasets grow in size and complexity, inference can quickly become prohibitively expensive, limiting the ambition of modelling in large-scale proteomics studies.


Traditional Markov Chain Monte Carlo (MCMC) methods such as Metropolis–Hastings \cite{mcmcmh}, Hamiltonian Monte Carlo (HMC) \cite{mcmchmc}, and the No U-Turn Sampler (NUTS) \cite{mcmcnuts}, are widely used to sample from posterior distributions. While effective, they are computationally expensive and depend on the convergence of a single chain, making them inherently sequential and hard to parallelise. This limits the complexity of models feasible in time-sensitive or resource-constrained settings. In public-health applications such as COVID-19 modelling \cite{Flaxman2020}, for example, computation has constrained the use of richly stratified models. Recent work including, hybrid MCMC \cite{Salimans2014MarkovCM}, many-short-chain MCMC methods \cite{wang2023taddaa}, and hybrid SMC for state-space models \cite{hirt2019scalable}, underscores the growing interest in scalable Bayesian inference.

SMC samplers constitute a naturally parallelisable alternative, particularly suited to modern multi-core and multi-thread computing environments. These population-based samplers approximate posterior distributions using weighted samples across multiple interacting particles \cite{smcsamplers}, and have demonstrated excellent scaling on distributed-memory, shared-memory, and GPU-based systems \cite{mpismc,gpusmc}. However, existing implementations typically rely on traditional high-performance computing (HPC) infrastructure, which is costly to acquire and operate, and often requires specialist expertise to use effectively. As a result, the benefits of SMC remain inaccessible to many practitioners, particularly in settings without guaranteed access to HPC resources.

To address this limitation, we introduce a novel opportunistic computing framework for distributed SMC samplers. Opportunistic computing harnesses underutilised resources, such as idle workstations, lab machines, or even mobile devices \cite{wcgdrugdiscovery,openzika}, to enable large-scale computation without the need for centralised or dedicated HPC infrastructure. Our framework builds on the HTCondor system and introduces a Coordinator-Manager-Follower (CMF) architecture, an extension of the Coordinator-Follower (CF) pattern, to enable scalable, fault-tolerant execution of SMC samplers in unpredictable and heterogeneous computing environments. A lightweight message-passing strategy ensures minimal overhead and adaptive load balancing, even under intermittent availability.

We demonstrate the effectiveness of our framework on a realistic quantitative proteomics model for shared peptide analysis, showing how opportunistic SMC enables scalable Bayesian inference without the infrastructural and financial costs of traditional HPC. This allows researchers in computational biology and related fields to adopt more ambitious models and broaden access to advanced inference techniques. To support adoption, we release CondorSMC, an open-source package for deploying opportunistic SMC samplers via HTCondor. It supports models written in both Python and Stan \cite{stanppl}, and is agnostic to model specification, making it easy to integrate into diverse scientific workflows.

\section{Quantitative Proteomics}
Quantitative proteomics plays a vital role in understanding complex diseases such as Alzheimer’s. By analysing protein expression across multiple brain regions, researchers can identify signatures of disease progression, regional vulnerability, and potential protective mechanisms, which are often missed when studying gene expression alone. For example, \cite{proteinexpression} conducted a spatially resolved proteomic study across six brain regions in Alzheimer’s patients and controls, quantifying over 5,000 proteins. The analysis revealed region-specific disruptions in immune, apoptotic, and metabolic pathways. These protein-level signatures offer valuable insights into disease progression and highlight potential avenues for early diagnosis and therapeutic intervention.

To extract clinically meaningful insights from proteomics data, researchers often rely on Bayesian models that account for uncertainty, noisy peptide measurements, and the ambiguity introduced by shared peptides (those originating from multiple proteoforms). While these models are highly informative, their application to large datasets presents a major computational challenge. Real-world proteomics experiments often involve thousands of peptides and hundreds of proteoforms, leading to high-dimensional parameter spaces and long inference times, even for state-of-the-art samplers. This limits the scalability of existing methods and restricts the use of advanced statistical models in settings where researchers need to compare multiple conditions, evaluate model variants, or conduct cohort-level analyses.

To address these challenges, we build on the shared-peptide model introduced by \citet{proteinquantification}, and applied in \citet{proteinexpression}, by applying our opportunistic SMC framework to estimate proteoform abundances from peptide-level data. The model explicitly captures the ambiguity of shared peptides by representing the mapping between proteoforms and their associated peptides. Implemented in Stan, it comprises 922 parameters, including proteoform-level abundances, assay effects, and peptide-specific noise terms. For demonstration, we use a synthetically generated dataset of 2,000 peptides mapped to a small number of proteoforms. By enabling scalable Bayesian inference on a realistic and representative model, our framework helps overcome the computational constraints that often limit the use of advanced modelling techniques across numerous scientific domains.

\section{Opportunistic Computing}
\label{sec:opportunistic_computing}
Commodity hardware is becoming increasingly powerful and is often underutilised under standard usage patterns. For instance, universities maintain large fleets of workstations that are actively used during the day by staff and students, but often remain idle at night and on weekends. Opportunistic computing is a form of distributed computing that capitalises on such idle resources, providing a cost-effective and accessible alternative to traditional HPC. Moreover, because it repurposes machines that are already powered on, opportunistic computing can offer improved energy efficiency.

Frameworks such as HTCondor \cite{htcondor} and BOINC \cite{boinc} are widely used to distribute computational workloads across idle computational resources. These systems have been successfully applied across a broad range of scientific domains. For example, HTCondor supported LIGO’s search for gravitational waves \cite{FAJARDO2021100679}, while BOINC enabled large-scale protein folding simulations through Folding@home \cite{wcgfoldingathome} and supported drug discovery efforts in the OpenZika project \cite{openzika}. Crowdsourced computing platforms have also contributed to accelerating scientific discovery in fields such as computational chemistry \cite{wcgdrugdiscovery}. At their peak, these distributed systems have rivalled the world’s most powerful supercomputers, for instance, Folding@home reached a peak performance of 1.5 exaFLOPS in March 2020.

In this work, we focus on HTCondor as a framework for distributing computational workloads. The University of Liverpool has an existing HTCondor pool comprising approximately 500 workstations, each equipped with a 6-core Intel i5 CPU and 16 GB of RAM. At peak availability, this cluster provides nearly 2,000 cores for computation. It is already used across disciplines ranging from medical and biological sciences to mathematics and engineering. An HTCondor pool consists of a central manager (or “scheduler”) daemon and multiple executor daemons. The central manager handles job distribution and monitoring, while executor nodes run the assigned tasks. HTCondor prioritises local users, meaning nodes are only allocated to distributed tasks when idle.

\begin{figure*}[t]
    \centering
    \includegraphics[width=\textwidth]{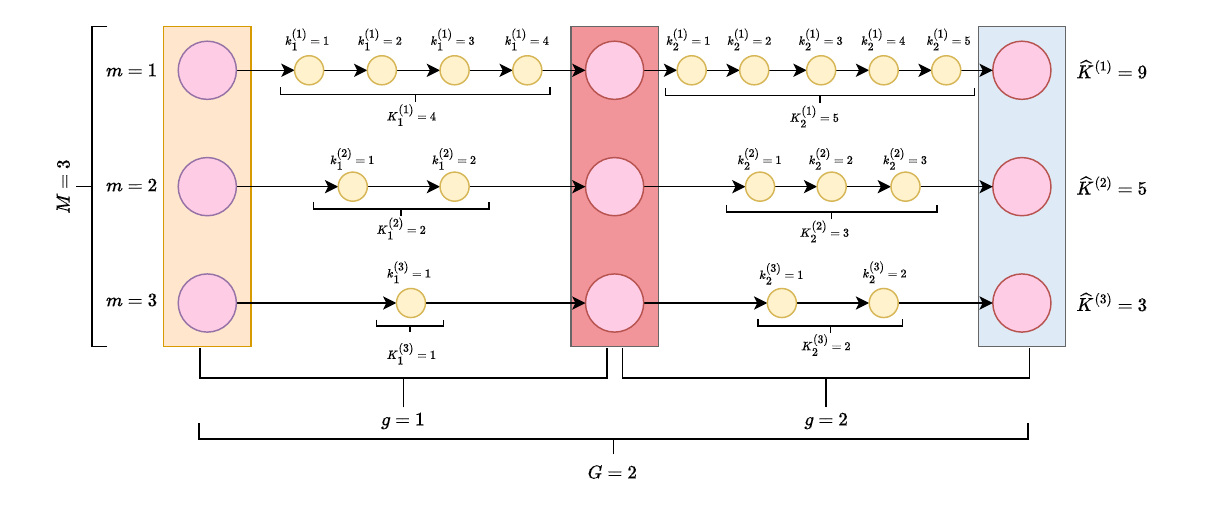}
    \caption{The evolution of an opportunistic SMC sampler.}
    \label{fig:chapter6:opportunisticsmcsampler}
\end{figure*}

\section{Opportunistic Sequential Monte Carlo Samplers}
\label{sec:opportunistic_SMC}
Readers are referred to the appendix for the necessary background on Bayesian inference and SMC samplers. The opportunistic SMC sampler extends the ``particle island model'' presented in \cite{paper:chapter6:particleisland}. The $N$ samples are divided among the $M$ nodes in the network such that the $m$th node handles $N^m = \left\lfloor \frac{N}{M} \right\rfloor$ samples, with the first $N \bmod M$ nodes each handling one additional sample. Rather than propagating samples for a fixed number of iterations, $K$, the nodes instead propagate samples for a number of global iterations, $G$. At each global iteration, nodes are requested to propagate their samples for a fixed amount of time, $T_g$.

This relaxation allows the sampler to naturally balance the load across the network, as nodes that are faster will perform more sampling iterations than the slower nodes. As a result, each node, $m$, performs a varying number of local iterations, $K^{m}_{g}$, at each global iteration. Each node performs a total of $\widehat{K}^{m} = \sum_{g=1}^{G} K^{m}_{g}$ over the $G$ global iterations of the sampler. Since nodes now propagate their samples over multiple iterations, they are susceptible to particle degeneracy. To mitigate this, nodes resample their local particle set when their local effective sample size (ESS) falls below a threshold (say, $\frac{N^m}{2}$).

To consistently index samples across varying local iterations in each global iteration, we define a function $\kappa(g, k, m)$ that maps node $m$’s local iteration $k$ in global iteration $g$ to a unique global index:

\begin{equation}
\label{eq:iterationfunc}
\kappa(g, k, m) = \left(\sum_{g^\prime=1}^{g-1} K^{m}_{g^\prime}\right) + k,
\end{equation}

\noindent
with $K^m_{g^\prime}$ the number of local iterations by node $m$ in global round $g^\prime$. We denote the resulting sample as $\textbf{x}^{(i, m)}_{\kappa}$, applying the same convention to related quantities.

In the opportunistic SMC sampler framework, the joint distribution across all states at the final iteration for the $m$th node is:

\begin{equation}
    \pi^{m}_{1:G}(\textbf{x}^{m}_{1:G}) = \pi_{G}(\textbf{x}_G^m)\prod^{G - 1}_{g=1}\prod^{K^{m}_{g}}_{k^{m}_g=1}\mathcal{L}\left(\textbf{x}^{m}_{\kappa-1} \middle| \textbf{x}^{m}_{\kappa}\right),
\end{equation}

\noindent
where $\textbf{x}^{m}_{\kappa}$ is the state at node $m$, at local iteration $k$ within global iteration $g$ and $\mathcal{L}\left(\textbf{x}^{m}_{\kappa-1} \middle| \textbf{x}^{m}_{\kappa}\right)$ is the local backward kernel at node $m$.

At the first iteration of the opportunistic SMC sampler, each node independently draws a set of $N^{m}$ samples from the initial proposal distribution $\{\textbf{x}^{(i, m)}_{1}\}_{i=1}^{N^{m}} \sim q_1(\textbf{x}_1^{(m)})$. Each node then calculates the initial importance weights for their sample set, $\{w^{(i,m)}_1\}_{i=1}^{N}$. The $i$th importance weight is:

\begin{equation}
    w^{(i, m)}_1 = \frac{\pi_1\left(\textbf{x}_1^{(i,m)}\right)}{q_1\left(\textbf{x}_1^{(i,m)}\right)}.
\end{equation}

In the following iterations, each node proposes a new set of samples $\{\textbf{x}_{\hat{k}^{m}}^{(i,m)}\}_{i=1}^{N}$ that is drawn from the previous set of samples $\{\textbf{x}_{\hat{k}^{m}-1}^{(i,m)}\}_{i=1}^{N}$ using a proposal distribution such that:

\begin{equation}
    q^{m}_{1:G}(\textbf{x}^{m}_{1:G}) = q_1(\textbf{x}_1^m)\prod^{G}_{g=1}\prod^{K^{m}_g}_{k^{m}_g=2} q\left(\textbf{x}^{m}_{\kappa} \middle| \textbf{x}^{m}_{\kappa-1}\right),
\end{equation}

\noindent
where $q(\textbf{x}_1^{m})$ is the initial proposal for the $m$th node and $q\left(\textbf{x}^{m}_{\kappa} \middle| \textbf{x}^{m}_{\kappa-1}\right)$ is the proposal distribution at the local iteration $k^m_g$ of global iteration $g$ within node $m$.

At iterations $\kappa^{m} > 1$, the unnormalised importance weights are calculated as:

\begin{equation}
    w^{(i, m)}_{\kappa} = w^{(i, m)}_{\kappa-1} \frac{\pi_k(\textbf{x}^{(i,m)}_{\kappa})}{\pi_{k-1}(\textbf{x}^{(i,m)}_{\kappa-1})} \frac{\mathcal{L}\left(\textbf{x}^{m}_{\kappa-1} \middle| \textbf{x}^{m}_{\kappa}\right)}{q\left(\textbf{x}^{m}_{\kappa} \middle| \textbf{x}^{m}_{\kappa-1}\right)}.
\end{equation}

At the end of each global iteration, the coordinator can pool the samples from the nodes and estimate expectations of a function using the global set of weighted samples:

\begin{align}
\mathbb{E}_{\pi(\textbf{x}_G)}[f(\textbf{x}_G)]
&= \int f(\textbf{x}_G)\,\pi(\textbf{x}_G)\, d\textbf{x}_G \nonumber\\
\label{eq:smcexpectation}
&= \int f(\textbf{x}_G)\,\pi(\textbf{x}_{1:G})\, d\textbf{x}_{1:G} \\
&\approx \frac{1}{\sum_{m=1}^{M} N^{m}} \sum_{m=1}^{M} \sum_{i=1}^{N^{m}} w^{(i,m)}_{G}\, f(\textbf{x}^{(i,m)}_{G}) .\nonumber
\end{align}

Note that, the samples and their associated weights can be used to estimate statistics at each local iteration, $\kappa$. These estimates could then be recycled \cite{essrecycling} to make use of all of the samples generated.

During each local iteration, nodes compute the ESS of their local sample set and perform resampling when needed. In each global iteration, the coordinator aggregates samples from all nodes and estimates expectations of the target distribution using the global sample set, as defined in (\ref{eq:smcexpectation}). The coordinator also calculates the ESS of the global sample set and resamples if necessary. The evolution of an opportunistic SMC sampler with $M = 3$ nodes and $G = 2$ global iterations and a varying number of samples and local iterations per node is visualised in Figure~\ref{fig:chapter6:opportunisticsmcsampler}.

\section{Adaptation of Proposal Hyperparameters}
Throughout this paper, we configure the opportunistic SMC sampler to use NUTS as the proposal distribution and its reverse as the backward kernel (see \cite{smcnuts} for details). To ensure efficient exploration of the target distribution, the proposal hyperparameters, namely, the step size and mass matrix, must be carefully selected.

In standard MCMC frameworks such as Stan, these hyperparameters are tuned during a dedicated ``warm-up'' phase that coincides with the ``burn-in'' period. Stan employs a windowed adaptation scheme, beginning with short windows and gradually transitioning to longer ones, refining stable estimates of the step size and mass matrix. Each chain adapts independently, with no information sharing, leading to potential variability in convergence speed and final hyperparameter values.

In contrast, our SMC framework uses a population of samples that jointly estimate the proposal hyperparameters. Rather than adapting in isolation, samples share information, allowing the ensemble to form stable and representative estimates more quickly. Since each node operates on a large sample set, accurate estimates of the hyperparameters of the proposal are often available early in the adaptation process.

We adapt Stan’s warm-up methodology to our setting as follows. After drawing the initial set of samples, warm-up begins by individually tuning each sample’s step size to its location in the target space using Algorithm 4 of \cite{hoffman2014no}. The samples then proceed through a sequence of adaptation windows with varying lengths and tolerances. Within each window, every sample runs $K^{w}$ NUTS iterations and adapts its step size using dual averaging (Algorithm 6 of \cite{hoffman2014no}). A global step size estimate is obtained by aggregating across all samples. This process continues until the global estimate stabilises (which is measured by averaging over $W_{tol}$ iterations) or until a maximum window size $W_{max}$ is reached. After each window, we compute the inverse sample variance to obtain a diagonal estimate of the inverse mass matrix, which is then shared among all samples.

As in any SMC procedure, particle degeneracy is a concern during adaptation. To mitigate this, we perform resampling whenever the ESS falls below a predefined threshold. Once warm-up concludes, the final global estimates of the step size and mass matrix are fixed, and the warm-up samples are discarded, mirroring the MCMC approach.

In our configuration, each compute node performs adaptation independently using its local set of $N^{m}$ samples. However, unlike Stan’s isolated chains, our method enables information sharing across all samples within a node, facilitating fast and accurate hyperparameter estimation. This distinction is particularly valuable in distributed or opportunistic computing environments, where weaker nodes may receive fewer samples. While we use large enough sample sets per node to ensure stability, the framework can be extended to support global adaptation across nodes, further addressing heterogeneity.

\section{A Framework for Deploying Opportunistic Sequential Monte Carlo Samplers on HTCondor}
In this section, we describe the opportunistic SMC framework, which distributes an SMC sampler across a large number of compute nodes using either a CF architecture or a novel extension: the CMF architecture. We leverage the HTCondor distributed computing framework to request compute resources for the duration of sampling and use a shared MySQL database to coordinate daemons and store sampling results.

The University of Liverpool’s HTCondor pool is heterogeneous, comprising both Windows and Linux machines. Due to restrictions in the vanilla HTCondor universe, jobs must execute independently without direct inter-daemon communication. Instead, daemons communicate via a shared MySQL database, hosted on a dedicated server accessible to all nodes. This database tracks daemon status, schedules jobs, stores checkpoints, and sampling outputs.

When a session is launched, the coordinator creates an entry in the \texttt{session} table and submits jobs requesting $M_m$ managers and $M_f$ followers. As daemons initialise, they register in the \texttt{pool} table and begin adapting NUTS hyperparameters locally. Once ready, followers sample while managers and the coordinator assign them jobs. Results are written to the \texttt{results} table, and checkpoints are periodically stored in the \texttt{checkpoint} table. The coordinator schedules jobs to all nodes, while managers schedule only to their assigned followers.

To ensure robustness in opportunistic environments, where compute units may join and leave the network at any time, we incorporate fault-tolerance mechanisms. HTCondor’s job migration system enables recovery in the event of node failure. If a daemon fails or becomes unavailable, HTCondor can migrate the job to a new executor. Upon resuming, the new daemon queries the database for an existing checkpoint. If one is found, it restores its state and continues sampling; otherwise, it initialises a new set of samples. Daemons that fail do not contribute to the global estimate until they have been successfully restored.

When many nodes are active, global aggregation and resampling can become a bottleneck. To address this, we extend the standard CF architecture into a hierarchical CMF architecture. In CMF, the coordinator delegates work to a set of managers, each of which supervises a subset of followers. Followers sample and resample independently within a fixed time budget, while managers coordinate this local work and return aggregated estimates to the coordinator. The coordinator then combines these manager-level summaries to update the global state and assign the next round of tasks. This layered design reduces bottlenecks, supports scalable sampling, and enables the system to adapt efficiently to changes in node availability.

\section{Evaluation and Discussion}
We evaluate the proposed framework using the quantitative proteomics model described earlier. This section is structured as follows: we begin by describing the network configuration and SMC hyperparameters, then assess the convergence of the proposal adaptation scheme. We follow this with an evaluation of the sampler's accuracy, and conclude with an analysis of its adaptability and scalability across heterogeneous compute environments.

\subsection{Network Configuration}
The performance of the opportunistic framework is evaluated under both CF and CMF architectures. In the CF setup, up to $N_f = 200$ followers operate independently. In the CMF configuration, $N_m = 4$ managers each supervise up to $N_f = 50$ followers. Both architectures run for $T_c = 1800$ seconds, with followers executing in $T_f = 300$ second intervals. In the CMF setup, managers report to the coordinator every $T_m = 600$ seconds and are assigned to reliable compute nodes expected to remain active for the full session, while followers run on opportunistic nodes that may drop out at any time. Followers propagate $N_f = 512$ samples and perform up to $K^{j}_{f} = 100$ local iterations per global iteration.

\subsection{Experimental Results}
To evaluate the framework’s ability to infer model parameters, we compare posterior estimates from the opportunistic SMC sampler to those obtained using Stan’s NUTS. Specifically, we run four NUTS chains, each with 50{,}000 warm-up iterations followed by 50{,}000 sampling iterations. Posterior means and standard deviations are computed from the post-warm-up samples. Stan is executed on a high-specification machine (Intel Core i7-12700F CPU, 32\,GB RAM), where the slowest chain takes 775 seconds to warm up and 825 seconds to complete sampling (1,600 seconds in total). This setup significantly exceeds the hardware capabilities of the HTCondor pool used in our experiments.

After warm-up, Stan reports a mean step size of $0.022 \pm 0.005$, while the proposed SMC method estimates $0.0630$, yielding an absolute error of $0.042$ (squared error of $1.76 \times 10^{-3}$). For the diagonal elements of the mass matrix, the mean squared error is $2.32 \times 10^{-3}$. The relative absolute error across dimensions averages $0.066$ (median $0.053$; maximum $0.506$), indicating generally small discrepancies with a few outliers. Adaptation on each node took $851 \pm 308$s, comparable to the $775$s warm-up of Stan’s slowest chain despite lower-spec hardware; these times can be reduced by exploiting within-node parallelism during adaptation.

We assess parameter estimates after the warm-up phase, comparing global posterior means from SMC to those obtained via Stan. For both architectures, the sampler satisfies the consistency criterion for all parameters, with each estimate deviating by less than 0.25 standard deviations from Stan's estimate:

\begin{equation}
\left| \frac{\mu_{\text{SMC}} - \mu_{\text{MCMC}}}{\sigma_{\text{MCMC}}} \right| < 0.25.
\end{equation}

Table~\ref{tab:results} summarises the average, minimum, and maximum errors (expressed in units of standard deviation) for each architecture. While higher errors are observed during the early sampling iterations, they steadily decrease as the sampler progresses. Notably, the smallest errors occur in the final iterations, where all parameters meet the convergence criterion, with estimates well below 0.25 standard deviations. This confirms that the sampler accurately recovers parameter estimates consistent with those obtained via Stan.

\begin{table}[ht]
    \centering
    \begin{tabular}{@{}lll@{}}
        \toprule
        \textbf{Metric} & \textbf{CF} & \textbf{CMF} \\ \midrule
        Average Error & $0.094$ & $0.086$ \\
        Minimum Error & $1.000 \times 10^{-5}$ & $1.000 \times 10^{-5}$ \\
        Maximum Error & $1.289$ & $1.136$ \\ \bottomrule
    \end{tabular}
    \caption{Deviation (in standard deviation units) of global post--warm-up parameter estimates from Stan posterior means for each architecture.}
    \label{tab:results}
\end{table}

In the CF architecture, the sampler runs for $T_c$ seconds and completes $G_f = 6$ global iterations, with each follower performing $K_f^j$ local iterations per round. In CMF, the sampler runs for 3 global iterations, during which 5 manager rounds are executed, with followers performing variable numbers of local iterations in each. Initialisation times vary: managers in CMF require $501 \pm 54$ seconds, and followers across both architectures take $940 \pm 301$ seconds.

\begin{figure}[ht]
    \centering
    \begin{overpic}[        trim=10pt 10pt 0pt 0pt,
        clip,
        width=0.42\textwidth]{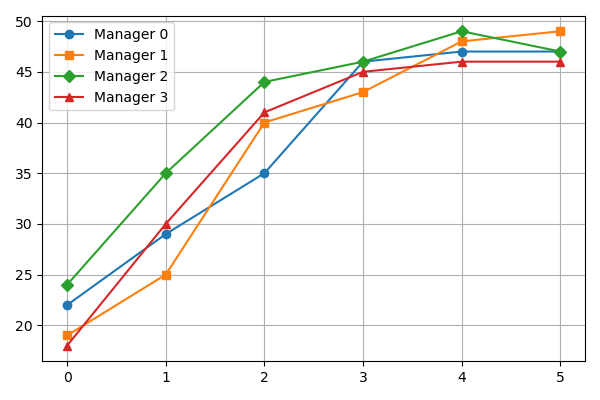}
        \put(32,-4){\scriptsize \textbf{Manager Sampling Iteration}}
        \put(-5,15){\rotatebox{90}{\scriptsize \textbf{Number of Active Followers}}}
    \end{overpic}
    \vspace{1em}
    \caption{Number of active followers per manager iteration in the CMF framework.}
    \label{fig:n_active_followers}
\end{figure}

Figure~\ref{fig:n_active_followers} illustrates how follower availability evolves over time during one run of the CMF architecture. Rather than all followers being present at the outset, new nodes join the sampling pool as they become available, resulting in a gradual increase in sampling capacity. The total number of active followers varies over time, reflecting the dynamic nature of the opportunistic compute environment. In periods of high external demand on the shared machine pool, follower participation becomes more volatile, leading to earlier or more frequent drop-offs. The observed stability (or instability) in follower availability is thus tightly coupled to background system utilisation.

\FloatBarrier
\FloatBarrier

\begin{figure}[ht]
    \centering
    \begin{overpic}[        trim=10pt 10pt 0pt 0pt,
        clip,
        width=0.42\textwidth]{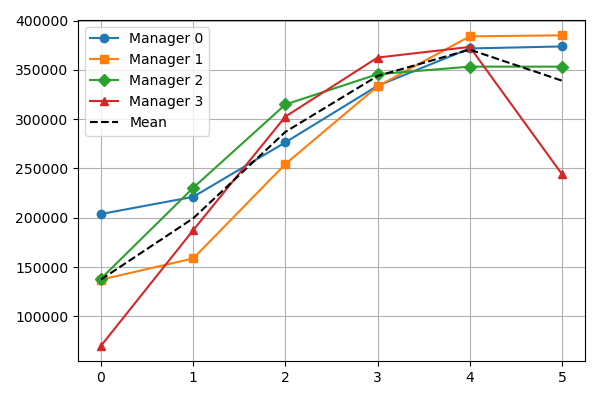}
        \put(35,-6){\scriptsize \textbf{Manager Sampling Iteration}}
        \put(-5,20){\rotatebox{90}{\scriptsize \textbf{Samples Generated}}}
    \end{overpic}
    \vspace{1em}
    \caption{Samples generated per global manager iteration by each manager’s follower set (fixed budget $T_m$). The mean across managers is also shown.}
    \label{fig:weak-scaling}
\end{figure}

As the sampler runs with a fixed time budget per global iteration, we evaluate weak scaling. Figure~\ref{fig:weak-scaling} shows that samples per iteration rise over successive manager iterations as more followers come online, so extra participants translate into higher throughput. Differences between managers, and the occasional drop, are due to uneven availability and workers joining or pausing. Overall, the mean increases, consistent with weak scaling. Throughput can be increased further by exploiting within-node parallelism.

\FloatBarrier
\FloatBarrier

\begin{figure}[ht]
    \centering
    \begin{overpic}[        trim=10pt 5pt 10pt 0pt,
        clip,
        width=0.4\textwidth]{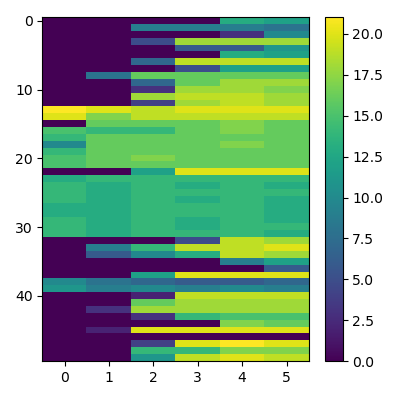}
        \put(26,-3){\small \textbf{Global Iteration ($g_f^m$)}}
        \put(-6,40){\rotatebox{90}{\small \textbf{Follower ID}}}
        \put(97,12){\rotatebox{90}{\small \textbf{Number of Local Iterations Performed ($K_f$)}}}
    \end{overpic}
    \vspace{1em}
    \caption{Number of local iterations ($K_f$) performed per global iteration ($g_f^m$) by followers allocated to manager three in a CMF network.}
    \label{fig:n_iters_per_follower}
\end{figure}

Figure~\ref{fig:n_iters_per_follower} illustrates the number of local sampling iterations performed by each follower at each manager iteration. Nodes contribute unevenly due to hardware heterogeneity and differences in when they join or leave the sampling pool. Our use of time-based iteration limits, rather than fixed iteration counts, enables natural load balancing: faster or longer-lived nodes perform more work. This design allows the framework to adapt to opportunistic environments, naturally balancing load and handling heterogeneity.

\FloatBarrier

\section{Conclusions and Future Work}
\label{sec:conclusions}
We present the first framework for distributing SMC samplers in opportunistic compute environments, enabling users to leverage these powerful methods without requiring access to expensive HPC infrastructure. The framework dynamically adapts to varying numbers of participating compute nodes, balances computational load across heterogeneous resources, and efficiently distributes tasks. This enables practitioners to create more ambitious models, resulting in better diagnostics at a potentially lower cost.

We demonstrated the system on a quantitative proteomics model implemented in Stan, selected for its practical relevance to domain experts. However, the framework itself is model-agnostic. The opportunistic SMC approach produced parameter estimates consistent with Stan’s built-in inference, and our adaptation of NUTS hyperparameters achieved comparable performance. Crucially, this adaptability enables our system to support the full spectrum of applications that Stan is used for. We encourage researchers to explore the potential of this framework to carry out ambitious, large-scale analyses that may have previously been impractical due to compute limitations.

While this first-of-its-kind system shows strong potential, several enhancements are planned. These include developing adaptive sampling methods to better support weaker nodes; scaling the database both horizontally and vertically to accommodate increased traffic and data volume; and, capitalising on within-node parallelism by vectorising core routines for adaptation, sampling, and resampling, thereby improving efficiency on multicore machines. Finally, we will apply the sampler to additional domains, including SMC$^{2}$ for epidemiological modelling \cite{rosato2023log}.


\section{Acknowledgments}
This work was funded by the EPSRC Centre for Doctoral Training in Distributed Algorithms at the University of Liverpool (EP/S023445/1) and IBM Research Europe. The authors would like to thank Dr Ian Smith from the University of Liverpool and Brian Ward from the Stan Development Team for their technical advice.

\setcounter{secnumdepth}{1}
\bibliography{aaai2026}

\appendix
\section{Numerical Bayesian Inference}
\subsection{Introduction}
This supplementary material provides concise technical background and fixes notation used throughout. Random vectors are bold: $\mathbf{x}\in\mathbb{R}^{D}$ (parameters/latents) and $\mathbf{y}\in\mathbb{R}^{D_N}$ (observations), where $D$ denotes the number of parameters and $D_N$ the total number of data dimensions.

The posterior follows Bayes' rule:

\begin{equation}
p(\mathbf{x}\mid\mathbf{y}) = \frac{p(\mathbf{y}\mid\mathbf{x})\,p(\mathbf{x})}{p(\mathbf{y})},
\end{equation}

The unnormalised target is written as:

\begin{equation}
\pi(\mathbf{x}) = p(\mathbf{x})\,p(\mathbf{y}\mid\mathbf{x}).
\end{equation}

Expectations with respect to a distribution $\pi$ are denoted $\mathbb{E}_{\pi}\!\left[f(\mathbf{x})\right]$.

For importance sampling (IS) and Sequential Monte Carlo (SMC) samplers, $q(\cdot)$ denotes a proposal and $w^{(i)}$ importance weights with normalised weights:

\begin{equation}
\tilde{w}^{(i)} = \frac{w^{(i)}}{\sum_{j=1}^{N} w^{(j)}}, \qquad
N_{\textrm{eff}} = \frac{1}{\sum_{i=1}^{N} \left(\tilde{w}^{(i)}\right)^2}.
\end{equation}

For SMC, we use intermediate targets $\{\pi(\mathbf{x})_{k}\}_{k=1}^{K}$, proposals $q(\mathbf{x}_{k}\mid \mathbf{x}_{k-1})$, and a backward L-kernel $\mathcal{L}(\mathbf{x}_{k-1}\mid \mathbf{x}_{k})$; particles are indexed by $i=1,\dots,N$ and iterations by $k=1,\dots,K$, with estimators:

\begin{equation}
\hat{f}_{k}=\sum_{i=1}^{N} \tilde{w}_{k}^{(i)} f(\mathbf{x}_{k}^{(i)}), \qquad
Z_{k}=\frac{1}{N}\sum_{i=1}^{N} w_{k}^{(i)}.
\end{equation}

For Markov Chain Monte Carlo (MCMC), $q(\mathbf{x}_k \mid \mathbf{x}_{k-1})$ denotes the proposal (transition) density and $\alpha(\mathbf{x}_{k-1},\mathbf{x}_k)$ the Metropolis–Hastings acceptance probability.

\subsection{Bayesian Inference}
Bayesian inference is based on Bayes' theorem, which updates the probability of a hypothesis (the parameter $\mathbf{x}$) given new evidence (the data $\mathbf{y}$). The framework requires a prior distribution $p(\mathbf{x})$, a likelihood $p(\mathbf{y}\mid \mathbf{x})$, and the marginal likelihood $p(\mathbf{y})$. The posterior distribution:

\begin{equation}
    p(\mathbf{x} \mid \mathbf{y}) = \frac{p(\mathbf{y} \mid \mathbf{x})\,p(\mathbf{x})}{p(\mathbf{y})},
\end{equation}

\noindent
is the primary object of interest, summarising uncertainty in $\mathbf{x}$ after observing $\mathbf{y}$, where:

\begin{equation}
    p(\mathbf{y}) = \int p(\mathbf{y} \mid \mathbf{x})\,p(\mathbf{x})\, d\mathbf{x}.
\end{equation}

In many models the marginal likelihood $p(\mathbf{y})$ is intractable, so we work with the posterior up to a normalising constant:

\begin{equation}
    p(\mathbf{x} \mid \mathbf{y}) \propto p(\mathbf{y} \mid \mathbf{x})\,p(\mathbf{x}).
\end{equation}

Hereafter, we denote the unnormalised target distribution by:

\begin{equation}
    \pi(\mathbf{x}) = p(\mathbf{x})\,p(\mathbf{y} \mid \mathbf{x}),
\end{equation}

\noindent
and its normalising constant by:

\begin{equation}
    Z = \int \pi(\mathbf{x})\, d\mathbf{x}.
\end{equation}

\subsection{Markov Chain Monte Carlo}
MCMC constructs a Markov chain $\{\mathbf{x}_k\}_{k=0}^{K}$ with stationary distribution $\pi(\mathbf{x}) \propto p(\mathbf{x})\,p(\mathbf{y}\mid \mathbf{x})$. We consider Metropolis–Hastings (MH) \cite{mcmcmh}, Hamiltonian Monte Carlo (HMC) \cite{mcmchmc}, and the No-U-Turn Sampler (NUTS) \cite{mcmcnuts}.

\paragraph{Metropolis–Hastings (MH).}
Given the current state $\mathbf{x}_{k-1}$, propose $\mathbf{x}_k \sim q(\mathbf{x}_k \mid \mathbf{x}_{k-1})$ and accept with probability:

\begin{equation}
\alpha\!\left(\mathbf{x}_{k-1},\mathbf{x}_k\right)
= 1 \wedge
\frac{\pi(\mathbf{x}_k)\, q(\mathbf{x}_{k-1}\mid \mathbf{x}_k)}{\pi(\mathbf{x}_{k-1})\, q(\mathbf{x}_k\mid \mathbf{x}_{k-1})}.
\end{equation}

If accepted, keep $\mathbf{x}_k$; otherwise set $\mathbf{x}_k=\mathbf{x}_{k-1}$. For symmetric proposals (e.g., random walk):

\begin{equation}
\alpha\!\left(\mathbf{x}_{k-1},\mathbf{x}_k\right)= 1 \wedge \frac{\pi(\mathbf{x}_k)}{\pi(\mathbf{x}_{k-1})}.
\end{equation}

The initial samples, known as the burn-in phase, are drawn before the sampler has fully converged and may be biased by initialisation. Discarding these ensures that posterior estimates reflect only well-mixed samples from the target distribution. After removing a burn-in of $\tau$ iterations, posterior expectations are estimated by:

\begin{equation}
\hat{f} = \frac{1}{K-\tau}\sum_{k=\tau+1}^{K} f(\mathbf{x}_k).
\end{equation}

\paragraph{Limitations of MH.}
Random–walk behaviour can lead to slow mixing in high dimensions and when $\pi(\mathbf{x})$ exhibits strong correlations or narrow ridges; performance is sensitive to proposal step size and often scales poorly with $D$.

\paragraph{HMC and NUTS.}
HMC augments the state with momentum $\mathbf{p}\sim\mathcal{N}(\mathbf{0},\mathbf{M})$ where $\mathbf{M}$ is referred to as the ``mass matrix''.

The Hamiltonian is defined as:

\begin{equation}
H(\mathbf{x},\mathbf{p}) = -\log \pi(\mathbf{x}) + \tfrac{1}{2}\,\mathbf{p}^{\top}\mathbf{M}^{-1}\mathbf{p}.
\end{equation}

At each iteration of HMC, a fresh momentum is drawn and the previous state $(\mathbf{x}_{k-1},\mathbf{p}_{k-1})$ is perturbed using the leapfrog integrator with $L$ steps of size $\epsilon$ to obtain $(\mathbf{x}_k,\mathbf{p}_k)$, accepted with probability:

\begin{equation}
\alpha = 1 \wedge \exp\!\big(-H(\mathbf{x}_k,\mathbf{p}_k)+H(\mathbf{x}_{k-1},\mathbf{p}_{k-1})\big).
\end{equation}

By integrating Hamiltonian dynamics, HMC suppresses random–walk behaviour and typically yields higher effective sample sizes than MH.

NUTS \cite{mcmcnuts} removes the need to pre-specify $L$ by extending trajectories forward and backward until a ``U-turn'' criterion is met, while adapting the step size and mass matrix during warm-up. This alleviates the main tuning burden of HMC and improves robustness on challenging posteriors.

\subsection{Importance Sampling}
IS is a Monte Carlo method for estimating expectations with respect to a target distribution using samples (particles) drawn from a different distribution \cite{book:chapter2:stochasticsimulation}. IS is typically used when the target is difficult to sample from directly, or to reduce the variance of a Monte Carlo estimator.

Given a function $f(\mathbf{x})$ and target $\pi(\mathbf{x})$, draw $N$ samples $\{\mathbf{x}^{(i)}\}_{i=1}^{N}$ from a proposal $q(\mathbf{x})$ with $q(\mathbf{x})>0$ wherever $\pi(\mathbf{x})>0$. Define importance weights $w^{(i)}=\frac{\pi(\mathbf{x}^{(i)})}{q(\mathbf{x}^{(i)})}$. Then:

\begin{align}
\label{eq:isexpectation}
\mathbb{E}_{\pi(\mathbf{x})}[f(\mathbf{x})]
&= \int f(\mathbf{x})\,\pi(\mathbf{x})\,d\mathbf{x}, \\
&= \int f(\mathbf{x})\,\underbrace{\frac{\pi(\mathbf{x})}{q(\mathbf{x})}}_{w(\mathbf{x})}\,q(\mathbf{x})\,d\mathbf{x}, \\
&\approx \frac{1}{N}\sum_{i=1}^{N} w^{(i)} f(\mathbf{x}^{(i)}).
\end{align}

For posterior expectations, $p(\mathbf{x}\mid\mathbf{y})$, we can work with the unnormalised target $\pi(\mathbf{x})=p(\mathbf{x})\,p(\mathbf{y}\mid\mathbf{x})$:
\begin{align}
    \mathbb{E}_{p(\mathbf{x}\mid\mathbf{y})}[f(\mathbf{x})]
    &= \frac{\int f(\mathbf{x})\,p(\mathbf{x})\,p(\mathbf{y}\mid\mathbf{x})\,d\mathbf{x}}
    {\int p(\mathbf{x})\,p(\mathbf{y}\mid\mathbf{x})\,d\mathbf{x}}\\
    &\approx \hat{f}
    = \frac{\sum_{i=1}^{N} w^{(i)} f(\mathbf{x}^{(i)})}{\sum_{i=1}^{N} w^{(i)}}\\
    &= \sum_{i=1}^{N} \tilde{w}^{(i)} f(\mathbf{x}^{(i)}),
\end{align}

where $\tilde{w}^{(i)}=\frac{w^{(i)}}{\sum_{j=1}^{N} w^{(j)}}$ are the normalised importance weights. This normalisation uses the fact that the sum of unnormalised weights estimates the normalising constant:

\begin{align}
p(\mathbf{y}) \approx Z = \frac{1}{N}\sum_{i=1}^{N} w^{(i)}.
\end{align}

When using IS, it is important that the proposal $q(\mathbf{x})$ place sufficient mass in the tails of $\pi(\mathbf{x})$ (i.e., be heavier-tailed) to control the variance of $\frac{\pi(\mathbf{x})}{q(\mathbf{x})}$. Otherwise, the importance weights can have very high (or infinite) variance, leading to poor approximations.

\subsection{Sequential Monte Carlo Samplers}
\label{sec:SMC_samplers}
Sequential Monte Carlo (SMC) samplers \cite{smcsamplers} combine importance sampling and resampling to provide unbiased estimates of the expectation of a function with respect to a target distribution, $\pi(\mathbf{x})$, where $\mathbf{x} \in \mathbb{R}^{D}$. Rather than targeting the distribution of interest directly, SMC samplers transition through a series of intermediate distributions $\{ \pi(\mathbf{x})_{k} \}_{k=1}^{K}$ such that the joint distribution of all the previous states at the final iteration $K$ is the target distribution of interest:

\begin{equation}\label{eq:smcsamplerjoint}
    \pi(\mathbf{x}) = \pi(\mathbf{x}_{1:K}) = \pi(\mathbf{x}_{K}) \prod_{k=2}^{K} \mathcal{L}(\mathbf{x}_{k-1} | \mathbf{x}_{k}).
\end{equation}

\noindent where $\mathcal{L}(\mathbf{x}_{k-1} | \mathbf{x}_{k})$ is referred to as the L-kernel or backward kernel and denotes the probability of moving back from $\mathbf{x}_{k}$ to $\mathbf{x}_{k-1}$. Careful choice of the L-kernel can significantly improve the efficiency of the SMC sampler. A number of L-kernels have been proposed in literature, including: the approximately optimal L-kernel outlined in \cite{smcsamplers}; the reverse of the proposal distribution outlined in \cite{smchmc, smcnuts}; and, a Gaussian approximation of the optimal L-kernel outlined in \cite{gaussoptlkernel}.

The first iteration of the SMC sampler follows the standard importance sampling procedure. We begin by drawing a set of $N$ samples from a user-defined initial proposal distribution, $q(\mathbf{x}_{1})$, such that $\{ \mathbf{x}_{1} \}_{i=1}^{N} \sim q(\mathbf{x}_{1})$. We then calculate an importance weight for each sample, $\{ w_{1}^{(i)} \}_{i=1}^{N}$, where:

\begin{equation}
    w_{1}^{(i)} = \frac{\pi(\mathbf{x}_{1}^{(i)})}{q(\mathbf{x}_{1}^{(i)})}.
\end{equation}

In the following iterations, a new set of samples $\{\mathbf{x}_{k}\}_{i=1}^{N}$ are drawn from the previous set $\{\mathbf{x}_{k-1}\}_{i=1}^{N}$ using a proposal distribution $q(\mathbf{x}_{k} \mid \mathbf{x}_{k-1})$. The user is free to choose any valid proposal distribution; a common choice is an MCMC kernel such as Random Walk, Hamiltonian Monte Carlo \cite{smchmc}, or the No U-Turn Sampler \cite{smcnuts}. At each iteration, the unnormalised importance weights are calculated as:

\begin{equation}
    w_{k}^{(i)} = w_{k-1}^{(i)} \frac{\pi(\mathbf{x}_{k}^{(i)})}{\pi(\mathbf{x}_{k-1}^{(i)})} \frac{\mathcal{L}(\mathbf{x}_{k-1}^{(i)} | \mathbf{x}_{k}^{(i)})}{q(\mathbf{x}_{k}^{(i)} | \mathbf{x}_{k-1}^{(i)})}.
\end{equation}

The joint distribution across all iterations is defined (in~(\ref{eq:smcsamplerjoint})) such that 
the set of weighted samples can then be used to estimate expectations of a function, $f\left(\mathbf{x}\right)$, (e.g. estimate the statistical moments of the target density) as:

\begin{align}    
\mathbb{E}_{\pi(\mathbf{x}_{k})} [f(\mathbf{x}_{k})] =& \int f(\mathbf{x}_{k}) \pi(\mathbf{x}_{k}) d\mathbf{x}_{k}, \\ 
=& \int f(\mathbf{x}_{k}) \pi(\mathbf{x}_{1:k}) d\mathbf{x}_{1:k}, \\
\approx& \frac{1}{N} \sum_{i=1}^{N} w_{k}^{(i)} f(\mathbf{x}_{k}^{(i)}).
\end{align}

Furthermore, if we wish to estimate expectations with respect to a posterior, $p\left(\mathbf{x}_k|\mathbf{y}\right)$, we can do so in a way that only requires the calculation of the prior, $p\left(\mathbf{x}_k\right)$, and likelihood, $p\left(\mathbf{y}|\mathbf{x}_k\right)$, by defining $\pi\left(\mathbf{x}_k\right)=p\left(\mathbf{y},\mathbf{x}_k\right)=p\left(\mathbf{x}_k\right)p\left(\mathbf{y}|\mathbf{x}_k\right)$ since:

\begin{align}
\mathbb{E}_{p(\mathbf{x}_{k}|\mathbf{y})} [f(\mathbf{x}_{k})] =&
\int f(\mathbf{x}_{k}) p\left(\mathbf{y}|\mathbf{x}_k\right) d\mathbf{x}_{k}, 
\\=&
\frac{\int f(\mathbf{x}_{k}) p\left(\mathbf{x}_k\right)p\left(\mathbf{y}|\mathbf{x}_k\right) d\mathbf{x}_{1:k}}{\int  p\left(\mathbf{x}_k\right)p\left(\mathbf{y}|\mathbf{x}_k\right) d\mathbf{x}_{1:k}},
\\\approx& \hat{f}_k = \frac{\sum_{i=1}^{N} w_{k}^{(i)} f(\mathbf{x}_{k}^{(i)})}{\sum_{i=1}^{N} w_{k}^{(i)}},\\
&=\sum_{i=1}^N \tilde{w}_k^i f(\mathbf{x}_k^{(i)}),
\end{align}

\noindent where we refer to $\tilde{w}_k^i=\frac{w_k^{(i)}}{{\sum_{i=1}^{N} w_{k}^{(i)}}}$ as the $i$th normalised importance weight and we note that we have capitalised on the fact that the sum of the unnormalised weights is proportional to an approximation to the normalisation constant:
\begin{align}
   p(\mathbf{y})\approx Z_k=\frac{1}{N}\sum_{i=1}^Nw_k^{(i)} .
\end{align}

Over the $K$ iterations of the SMC sampler, the variance of the samples increases, such that only a subset retain high normalised importance weights. This ``degeneracy'' phenomenon can be mitigated by resampling the samples at each iteration. While alternatives exist, the simplest resampling variant involves independently sampling with replacement from the multinomial distribution defined by the normalised weights to generate a list of indexes. Each member of the new particle population is a copy of the member of the old population at the corresponding index. The new samples are all given an unnormalised weight of $\frac{Z_k}{N}$.

The effective sample size (ESS) is a measure of the number of samples that have a high importance weight associated with them where
\begin{align}
    N_{\textrm{eff}} = \frac{1}{\sum_{i=1}^N \left(\tilde{w}_k^{(i)}\right)^2}.
\end{align}
Resampling is undertaken if the $N_{\textrm{eff}}$ falls below some pre-defined threshold, usually $\frac{N}{2}$.

After sampling is complete, particle recycling can be used to calculate estimates of expectations pertinent to the target distribution in a way that uses samples generated at all the iterations of the SMC sampler (and not just the final iteration). To generate the combined estimate, $\tilde{f}$, each iteration is given a weight, where the weight of the $k$th iteration is $\lambda_k$:

\begin{equation}
    \tilde{f} = \sum_{j=1}^{K} \lambda_{j} \hat{f}_{j}.
\end{equation}

This weight, $\lambda_k$, can be defined in a number of ways, such as using ESS or the approximation to the normalising constant: the reader interested in further discussion about these choices is referred to~\cite{essrecycling}.

\end{document}